\documentclass[onecolumn,12pt]{revtex4-2}
\usepackage{graphicx}
\usepackage{bm}
\usepackage{color,ulem,amssymb,amsmath}
\usepackage{mathrsfs}

\newcommand{\de}{\mathrm{d}}

\begin{document}

\title[Effective modes in FPUT]{A mean-field theory of effective normal modes in the Fermi-Pasta-Ulam-Tsingou model}

\author{Antonio Ponno}
\address{Department  of  Mathematics  ``T.  Levi-Civita'',  Universit\`a di  Padova,  Via  Trieste  63,  35131  Padova,  Italy}
\email{ponno@math.unipd.it}
\author{Giacomo Gradenigo}
\address{Dipartimento di Fisica e Astronomia ``Galileo Galilei'', Universit\`a di Padova, Via Marzolo 8, 35131 Padova, Italy}
\address{Gran  Sasso  Science  Institute,  Viale  F.  Crispi  7,  67100  L'Aquila,  Italy}
\address{INFN-LNGS, Via G. Acitelli 22, 67100 Assergi (AQ), Italy}
\email{giacomo.gradenigo@unipd.it}
\author{Marco Baldovin}
\address{Istituto dei Sistemi Complessi - CNR, P.le A. Moro 5, I-00185, Rome, Italy}
\email{marco.baldovin@cnr.it}
\author{Angelo Vulpiani}
\address{Dipartimento  di  Fisica,  Universit\`a  di  Roma  ``Sapienza'',  P.le  A.  Moro  5,  I-00185,  Rome,  Italy\\
  Istituto dei Sistemi Complessi - CNR, P.le A. Moro 5, I-00185, Rome, Italy}
\email{angelo.vulpiani@roma1.infn.it}


\begin{abstract}
  We present a non-perturbative, mean-field theory for the
  Fermi-Pasta-Ulam-Tsingou model with quartic interaction, capturing
  the {quasiperiodic} features shown by the system at all energies in
  the thermodynamic limit. Starting from the true Hamiltonian $H$ of
  the system with $N$ degrees of freedom, we introduce a {\it
    mean-field Hamiltonian} $\mathcal{H}$ such that the difference
  $h_N=(H-\mathcal{H})/N$, considered as a random variable with
  respect to the Gibbs measure, tends to zero as $N\to\infty$, in
  probabilistic sense. The dynamics of the mean-field Hamiltonian
  $\mathcal{H}$ consists of $N$ independent oscillation modes with
  renormalized frequencies $\Omega_k =
  \omega_k\sqrt{1+\gamma(\varepsilon)}$, $\omega_k$ being the
  frequency of the $k$-th normal mode of the linearized system,
  whereas $\gamma(\varepsilon)$ is an explicit function of the
  specific energy $\varepsilon$ of the system. Analytical predictions
  drawn from the effective Langevin equations ruling the dynamics of
  such oscillation modes are successfully compared with the numerical
  data from the original Hamiltonian dynamics.  Such a simple
  decomposition of the true dynamics into $N$ {\it effective normal
    modes} holds at all energy scales, i.e. from the quasi-integrable
  regime to the strongly chaotic one.
\end{abstract}

%
%
%
%
%


\maketitle

\section{Introduction}
\label{sec-0}

One among the most important and longstanding debates about the
foundations of statistical mechanics concerns chaos -- specifically, the role it 
plays in the thermalization processes of classical many-body systems.
Investigation in this direction was triggered by the seminal work of
Fermi, Pasta, Ulam and Tsingou~\cite{FPUT55,Tsin}. In this seminal contribution, the authors
noticed that{, at variance with the expectations based on the Poincaré-Fermi theorem,} a small non-linear correction to the interactions of an harmonic chain  was not sufficient to
produce equipartition among all the degrees of freedom, within the
observational time-scale, when the model was prepared in a very atypical
initial condition. Such a result, named after the authors the FPUT problem, dictated the line of research for the following decades: how far from its integrable limit a model must be
in the space of parameters and/or initial conditions in order to observe thermalization?  A
question which can also be posed in the following terms: which is the
amount of {nonlinearity} needed to test good thermalization
properties? Such an approach to the dynamics of non-linear systems clearly
follows the point of view according to which the presence of strong
chaos is a sufficient condition to guarantee the validity of
statistical mechanics. What we want to stress here is that this
assumption is, at least, doubtful. There are counter-examples showing
that, even in the presence of positive Lyapunov exponents, remarkable
deviations from the expected equilibrium behaviour can be found: this
is for instance the case for the absence of thermalization for the
highly chaotic phase of rotators discussed in~\cite{LPRV87}, or for
the manifest lack of self-averaging and presence of fat tail in the
distribution of Lyapunov exponents for a system of coupled chaotic
maps found in~\cite{FMV91}. We must therefore acknowledge the
evidences that not only the presence of chaos does not guarantee
thermalization~\cite{CFLV08,BGV21}, but, furthermore, even the lack of
chaos does not prevent thermalization to be detected, a result which
crucially depends on which variables/observables are taken into
account. There are examples showing that, by considering the
appropriate observables, a kind of fast thermalization, starting from atypical initial
conditions, can be observed even in integrable
systems~\cite{BVG21,CVG22, BMV23, CBLMV24}.

Having assessed the fact that chaos and thermalization are not
necessarily in a cause-and-effect relationship, it seems then
legitimate and interesting to ask ourselves a further question: is it
possible that chaotic systems display, at some level of description,
regular properties which we may refer to, in a broad sense, as
``effective'' or ``emerging'' quasi-integrability?  Such a question
was first addressed, to our knowledge, in a series of papers by
Casartelli and co-workers concerning the behaviour of physical
quantities {\it conserved on average} along the Hamiltonian dynamics
of the FPUT model with quartic interaction. Their remarkable finding
was that the energy of individual Fourier modes, up to a
renormalization of the characteristic frequencies, is conserved on
average at {\it all} energy scales~\cite{ACM95,AC01,AC02}. More
precisely, they found that the FPUT model with $N-1$ degrees of
freedom displayed exactly $N-1$ quantities conserved on average along
the dynamics, namely the ``harmonic'' energies $\mathcal{E}_k$ defined
as
\begin{equation}
\label{Ekeff}
\mathcal{E}_k = \frac{1}{2}\left(P_k^2 + \Omega_k^2 Q_k^2\right)\ ,
\end{equation}
where $Q_k$ and $P_k$ are the canonical coordinates and momenta of the Fourier modes, with renormalized frequencies
\begin{equation}
\label{Omkpre} 
\Omega_k(\varepsilon)= \sqrt{1 + \gamma(\varepsilon)}\ \omega_k\ .
\end{equation}
Here $\omega_k$ is the characteristic frequency of the $k$-th normal mode of the linearized dynamics, whereas $\gamma(\varepsilon)$ is a suitable function, depending on the specific energy $\varepsilon$ of the system, determined numerically. Of course, $\gamma(\varepsilon)\to0$ as $\varepsilon\to0$, so that the linear integrable regime is recovered at low specific energies. On the other hand, the mentioned authors found a $\gamma$ growing with $\sqrt{\varepsilon}$ as $\varepsilon\to\infty$, putting into evidence the surprising robustness of the result up to the strongly chaotic regime completely dominated by the quartic interaction between the Fourier modes. The picture emerging from~\cite{ACM95,AC01,AC02} is that the non-linear FPUT chain turns out
to be fully equivalent, {\it at all energy scales} and up to thermal
fluctuations, to a set of uncoupled harmonic oscillators with renormalized
frequencies, named ``quasi-normal modes'' by those authors. 

The somehow overlooked results of Casartelli and co-workers
really represent, to us, a change of perspective in the study of FPUT-like
models.  The standard approach up to now consisted indeed in studying the weakly
non-linear regime and in trying to understand the long time-scales to thermalization 
in terms of closeness to and departure from integrability. In this respect, while
the FPUT model with quartic
non-linearity can be regarded as a perturbation of the harmonic oscillators, it is well known that the integrable model controlling the dynamics of the FPUT with cubic interaction, in the weak coupling regime, is the Toda model~\cite{Mana,FFML,PCSF, BCP13, BPP18, BOP}. The logic of the overwhelming number of
studies in the literature on the FPUT models is genuinely
perturbative~\cite{BI05,CGG05,G07}, consisting in finding a smart change of
variables allowing  to single out a close integrable
model, a paradigm which works well only in the weakly
nonlinear regime. This fact is not perceived as a limitation, 
since the strongly chaotic/nonlinear regime is somehow dubbed as 
equivalent to a stochastic dynamics for all FPUT-like models, and the quest
for interesting physics is concentrated at and below the so-called
{\it ``strong stochasticity threshold''}~\cite{LPRSV85,CLP95}. From such a point of view, 
the mentioned works of Casartelli and co-workers trigger a
remarkable shift of perspective, showing that interesting
physical phenomena may take place at all energy scales, no matter the regime, in a comforting unified picture.

Other relevant works on the way paved by Casartelli and co-workers are, so far, that  of 
Lepri~\cite{L98} and those from Lvov and co-workers~\cite{GLC05,GLC07}.  In~\cite{L98}
the author provides a derivation of an effective dynamics for
independent Fourier modes with renormalized frequencies by exploiting
the Zwanzig-Mori projection formalism and some reasonable closures for
the dynamics of multipoint correlation functions, being eventually
able to obtain effective Langevin-like equations for harmonic oscillators
with renormalized frequencies. A similar result is obtained
in~\cite{GLC05,GLC07} within a wave-turbulence approach where non
resonant interactions are neglected and the dynamics of the
renormalized Fourier modes is obtained again by means of
convenient approximations on the multimode correlation functions.
However, in both cases, the results are not completely well founded from a mathematical
point of view: apart from technical
details in the derivations, they both represent {\it bona-fide} expansions around the dominant processes at the
hydrodynamic scale.

The aim of the present work is to show that, for the FPUT model with quartic interaction, the decomposition of the dynamics into effective normal modes with renormalized frequencies can be partly proven by a new strategy, in the spirit of the
 approach to the ergodic {problem in Statistical Mechanics} introduced by
Khinchin~\cite{K49}, and extended by Mazur and Van Der Linden~\cite{MVDL}{, valid for systems with many degrees of freedom}. Here is a summary of our results.
\begin{itemize}
\item We identify a mean-field Hamiltonian $\mathcal{H}$ which is close, in measure, to the true Hamiltonian $H$ of the model. More precisely, 
 we are able to show that the normalized distance $h_N=(H - \mathcal{H})/N$, regarded as a random variable with respect to a suitable Gibbs measure, tends to zero
 in the limit $N\rightarrow\infty$, at almost any point in the phase space. 
 \item The Hamiltonian dynamics generated by $\mathcal{H}$ turns out to be, at all energy scales, and in the limit  $N\rightarrow\infty$, the linear dynamics of effective normal modes with the renormalized frequencies \eqref{Omkpre}. In particular, the renormalization coefficient $\gamma(\epsilon)$ appearing in \eqref{Omkpre} is analytically determined and shown to be in full agreement with the numerical results.
\item The actual dynamics of the single, effective normal mode is shown to be ruled by a damped-forced harmonic oscillator equation, with white noise forcing and suitable damping coefficient. The latter coefficient is shown to be in full agreement with the one previously determined by Lepri~\cite{L98} on the hydrodynamic scale.  
\end{itemize}

The paper is organized as follows: in Sec.~\ref{sec-2} we detail, comment and justify 
the main results of the paper; in Sec.~\ref{sec-3} we
present a description of the stochastic dynamics of the effective normal modes consistent with
the mean-field theory, and compare all the analytical predictions with
the numerical data from the original model; finally, Sec.~\ref{sec-4} is devoted to 
conclusions and comments. Further technical aspects are discussed in the Appendices.

\section{Model, outline and discussion of the main results}
\label{sec-2}

In this section we introduce the model and present the main results, discussing their meaning and giving a short justification of their derivation. Let
us stress again the non-perturbative approach of our
work. We do not seek a canonical change of variables making the
degrees of freedom of the system weakly coupled, as usual in Hamiltonian perturbation theory. We instead define another Hamiltonian $\mathcal{H}$ whose distance from the true one, regarded as a random variable with respect to the Gibbs distribution, is small in the thermodynamic limit, for any value of the specific energy, or temperature.  

\subsection{The FPUT quartic model and its properties}

Let us
recall the standard form of the FPUT model Hamiltonian with quartic
interactions between the modes, namely
\begin{equation}
   H =  K({\bf p}) + U_2({\bf q}) + g\, U_4({\bf q})
  \label{H}
\end{equation}
where the kinetic, quadratic and quartic potential energy terms are defined as follows
\begin{equation}
  K({\bf p})   =  \frac{1}{2} \sum_{n=1}^{N-1}   p_n^2\ ;
\label{K}
\end{equation}
\begin{equation}
  U_2({\bf q})  =  \frac{1}{2} \sum_{n=0}^{N-1} (q_{n+1}-q_n)^2\ ; 
\label{U2}
\end{equation}
\begin{equation}
  U_4({\bf q})  =  \frac{1}{4} \sum_{n=0}^{N-1} (q_{n+1}-q_n)^4\ . 
\label{U4}
\end{equation}
In the quantities above, as usual, the variables $q_i$ denotes the relative
displacement from the equilibrium position (zero force) of particle $i$, whereas $p_i$ is the corresponding conjugate momentum. We consider fixed boundary conditions: $q_0=q_N=p_0=p_N=0$ for $N-1$ moving particles. Finally, concerning the coupling constant $g>0$ in
\eqref{H}, it might be rescaled to one, but we prefer to keep it as a bookmark of the nonlinearity. 

One easily checks that in terms of the usual Fourier (or normal mode) canonical variables
\begin{equation}
\label{QkPk}
Q_k =\sqrt{ 2 \over N } \sum_{n=1}^{N-1} q_n  \sin \left( {n k \pi \over N } \right)\ \ ;\ \ 
P_k = \sqrt{ 2 \over N } \sum_{n=1}^{N-1} p_n  \sin \left( {n k \pi \over N } \right)\ , 
\end{equation}
the quadratic part of the Hamiltonian \eqref{H} reads
\begin{equation}
  K+U_2 =  \sum_{k=1}^{N-1} \frac{ P_k^2 +  \omega_k^2 Q_k^2}{2}\ ,
  \label{H2}
\end{equation}
where the frequencies $\omega_k$ are given by
\begin{equation}
\label{omegak}
\omega_k= 2 \sin \left( { k \pi \over 2N } \right)\ ;\ \ k=1,\dots,N-1\ .
\end{equation}
Thus, {for $g= 0$} the Hamiltonian \eqref{H} of the system reduces to \eqref{H2}, i.e. to the sum of $N-1$ non interacting harmonic oscillator Hamiltonians, so that the harmonic energies $E_k = \frac{1}{2} (P_k^2 +\omega_k^2 Q_k^2) $ are $N-1$ constants of motion in involution (pairwise vanishing Poisson bracket).
The system is thus integrable, its quasi-periodic dynamics consisting of $N-1$
independent harmonic oscillations, or normal modes, determined by the equations
\begin{equation}
\ddot{Q}_k= -\omega_k^2 Q_k\ .
\label{Qddotlin}
\end{equation}
This is a peculiar feature of the FPUT models with quartic or higher order {\it even} anharmonicity: in the weakly non-linear
regime they can be regarded as perturbations of uncoupled harmonic oscillators. On the other hand, as is well known, FPUT models with nonzero cubic anharmonicity, the ``underlying integrable dynamics'' in the weak coupling regime is ruled by the nonlinear, integrable Toda system~\cite{BCP13,BPP18}. 

\subsection{Main results: statistics}

As explained in the Introduction, the existence of {\it effective normal modes}, or independent ``harmonic'' oscillations, at all energy scales, and for typical initial data, has been put into evidence for the Hamiltonian system \eqref{H}.   
The main result of the present work is to explain such a simple decomposition of the dynamics of the system showing that the {\it mean-field Hamiltonian}
\begin{equation}
  \mathcal{H}({\bf q},{\bf p}) = K({\bf q},{\bf p}) + U_2({\bf q}) + \frac{\alpha(\varepsilon)}{N} [U_2({\bf q})]^2
  \label{Hmf}
\end{equation}
is close to the true Hamiltonian \eqref{H} in probabilistic sense. Inspired by the previous works \cite{ACM95,AC01,AC02}, we define the coefficient $\alpha(\varepsilon)$ in such a way that 
$\langle H \rangle_\mu=\langle \mathcal{H} \rangle_\mu$, where $\langle\cdot\rangle=\int\cdot\ \de\mu$ denotes the mean value with respect to the {\it canonical Gibbs measure conditioned to fixed ends}, namely
\begin{equation}
\label{condGibbs}
\de \mu({\bf q},{\bf p})=\frac{e^{-\beta H({\bf q},{\bf p})}\big|_{q_0=q_N=0}
\de q_1\dots\de q_{N-1}\de p_1\dots\de p_{N-1}}{
\int e^{-\beta H({\bf q},{\bf p})}\big|_{q_0=q_N=0}
\de q_1\dots\de q_{N-1}\de p_1\dots\de p_{N-1}}\ ,
\end{equation}
where $\beta=1/\theta$ denotes the canonical inverse temperature. Thus we set
\begin{equation}
\label{alphabeta}
\alpha(\beta)=g\frac{N\langle U_4\rangle_\mu}{\langle U_2^2\rangle_\mu}\ .
\end{equation}
Concerning the specific energy $\varepsilon$, it is defined as usual by
\begin{equation}
\label{epsbeta}
\varepsilon(\beta)=\frac{\langle H\rangle_\mu}{N}\ .
\end{equation}
What we refer to as $\alpha(\varepsilon)$ in the paper, with some abuse of notation, is the graph of the curve 
$\beta\mapsto(\alpha(\beta),\varepsilon(\beta))$, defined for all $\beta>0$.  
We are now in the position to state our main result.
{\it 
Let
\begin{equation}
\label{hN}
h_N({\bf q})\equiv\frac{H({\bf q},{\bf p})-\mathcal{H}({\bf q},{\bf p})}{N}=
\frac{NgU_4({\bf q})-\alpha(\varepsilon)[U_2({\bf q})]^2}{N^2}\ ,
\end{equation} 
be the normalized distance between $H$ and $\mathcal{H}$ at a point $({\bf q},{\bf p})$ of the phase space. Then 
\begin{equation}
\label{muweak}
\lim_{N\to\infty}\mu\left(\left\{({\bf q},{\bf p}):\ |h_N({\bf q})|>N^{-a}\right\}\right)=0\,,
\end{equation}
where  $\de\mu$ is the measure \eqref{condGibbs},   for any exponent $0<a<1/2$. {We can also prove that} the much stronger conclusion
\begin{equation}
\label{mustrong}
\mu\left(\left\{({\bf q},{\bf p}):\ \lim_{N\to\infty}h_N({\bf q})=0\right\}\right)=1
\end{equation}
actually holds. Moreover, 
\begin{equation}
\label{alphaeps}
\lim_{N\to\infty}\alpha(\beta)= g\frac{\langle r^4\rangle_\phi}{\langle r^2\rangle_\phi^2}\ \ ;\ \ 
\lim_{N\to\infty}\varepsilon(\beta)= \frac{1}{2\beta}+\langle \phi \rangle_\phi\ ,
\end{equation}
where $\phi(r)=r^2/2+gr^4/4$ is the pair potential and 
$\langle\cdot\rangle_\phi=\int(\cdot)e^{-\beta\phi(r)}\de r/\int e^{-\beta\phi(r)}\de r$ denotes the mean value with respect to the single spring measure.}

\medskip

\noindent
The detailed proof of \eqref{muweak}-\eqref{alphaeps} is reported in~\ref{Astat}. We now discuss a bit the meaning of the result and its implications, and outline the main ideas underlying the proof at the end of the present Section.

{
\subsection{Implications of the results}
}

\label{sec:results-implications}
The result \eqref{muweak} is an estimate {in the spirit of} Khinchin. Choosing for example $a=1/4$, \eqref{muweak} means that the probability that
$|H-\mathcal{H}|>N^{3/4}$ tends to zero {as $N\to\infty$ (or, equivalently, the probability that
$|H-\mathcal{H}|<N^{3/4}$ tends to one as $N\to\infty$)}. On the other hand, the stronger result \eqref{mustrong} - notice that \eqref{mustrong} implies \eqref{muweak} - means (see the proof)  that the difference $|H-\mathcal{H}|$ grows slower than $N^{3/4}$ at almost any point in the phase space. In this sense, the two Hamiltonians are close to each other. In a sense, one can interpret the result from the perspective of perturbation theory. Indeed, one can write the FPUT Hamiltonian \eqref{H} as 
\begin{equation}
H=\mathcal{H}+Nh_N\ ,
\end{equation}
being tempted to read $\mathcal{H}$ as a statistical normal form $O(N)$, and $Nh_N$ as a small remainder $O(N^{3/4})$. A natural question in this respect is whether in the limit $\varepsilon\to0$ the mean field Hamiltonian matches the first order Birkhoff normal form computed by the standard means of canonical perturbation theory (see e.g. \cite{Giorg}). The affirmative answer is proven in~\ref{Apert}.

\begin{figure}
    \centering
    \includegraphics[width=.7\linewidth]{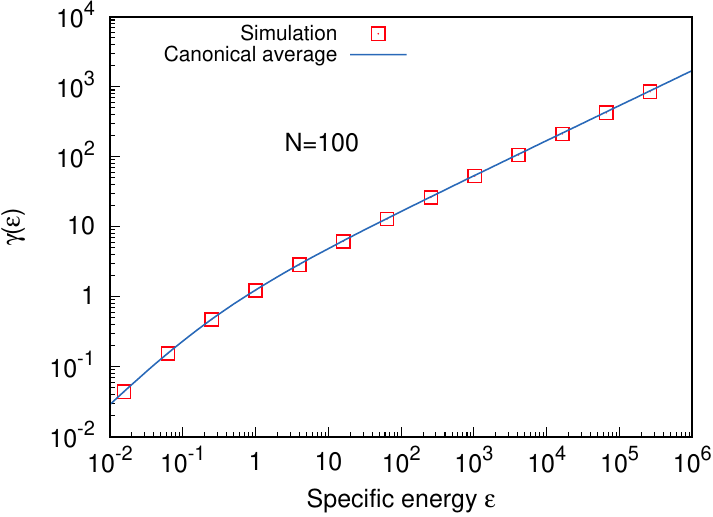}
    \caption{Renormalizing coefficient $\gamma(\varepsilon)$, as a function of the specific energy. The blue solid curve represents the (canonical) ensemble estimate on the r.h.s. of Eq.~\eqref{gamma}. Red squares are computed considering empirical averages in molecular dynamics simulations of model~\eqref{H}{, according to Eq.~\eqref{gammaemp}}. We use a symplectic integration algortihm (Verlet), with a time-step chosen, for each value of $\varepsilon$, in such a way that the relative fluctuations on the total energy due to numerical effects are $\sim O(10^{-5})$. Parameters: $N=100$, $g=1$.   }
    \label{fig:gamma}
\end{figure}

The remarkable feature of the mean-field Hamiltonian \eqref{Hmf} is that its associated dynamics is ``asymptotically free'' in the large-$N$ limit, i.e. it consists of
uncoupled harmonic oscillations with renormalized frequencies, i.e. frequencies proportional to those of the linearized model through an energy dependent factor. Indeed, when the Hamiltonian 
$\mathcal{H}$ is written in terms of the canonical variables \eqref{QkPk}, its Hamilton equations read
\begin{equation}
\ddot{Q}_k=-\frac{\partial\mathcal{H}}{\partial Q_k}=-\left(1+2\alpha(\varepsilon)\frac{U_2}{N}\right)
\frac{\partial U_2}{\partial Q_k}\ .
\end{equation}
Now, recalling that $U_2=\sum_{k=1}^{N-1}\omega_k^2Q_k^2/2$, one gets 
\begin{equation}
\label{Qkmf1}
\ddot{Q}_k=-\left(1+2\alpha(\varepsilon)U_2/N\right)\omega_k^2Q_k\ .
\end{equation}
Notice that $U_2$ is not a constant of motion, either for the true dynamics and for the mean field one. Thus, strictly speaking, \eqref{Qkmf1} does not define a simple harmonic dynamics. However, 
recalling the definition \eqref{U2} of $U_2$, one can invoke for $U_2/N$ a law of large numbers and substitute it by its mean value $\langle U_2\rangle_\mu/N$ {in the interacting system. This can be formally proved showing that}
\begin{equation}
\label{U2law}
\mu\left(\left\{({\bf q},{\bf p}):\ \lim_{N\to\infty}\frac{U_2({\bf q})-\langle U_2\rangle_\mu}{N}=0\right\}\right)=1\ ;
\end{equation}
\begin{equation}
\label{U2lim}
\lim_{N\to\infty}\frac{\langle U_2\rangle_\mu}{N}=\frac{\langle r^2\rangle_\phi}{2}\ ,
\end{equation}
{where $d \mu$ is the probability measure in the interacting
  system; the} proof is also reported in~\ref{Astat}. Taking into
account the latter relations, one can replace $U_2/N$ with its
expectation on the right hand side of \eqref{Qkmf1}, to obtain
\begin{equation}
\label{Qkmf2}
\ddot{Q}_k =-\Omega_k^2(\varepsilon)Q_k\ .
\end{equation}
Here the renormalized frequencies $\Omega_k$, in the large $N$ limit, are given by
\begin{equation}
\label{Omk}
\Omega_k=\sqrt{1+\gamma(\varepsilon)}\ \omega_k\ ,
\end{equation}
where the coefficient $\gamma(\varepsilon)$, taking into account \eqref{Qkmf1} and \eqref{U2law}, is defined as follows:
\begin{equation}
\label{gamma}
\gamma(\varepsilon) {\equiv}2\alpha(\varepsilon)\frac{\langle U_2\rangle_\mu}{N}
\sim g\frac{\langle r^4\rangle_\phi}{\langle r^2\rangle_\phi}\ ,
\end{equation}
the asymptotic value being valid for large $N$ as a consequence of \eqref{alphaeps} and \eqref{U2lim}.
{The demonstration that, at all energy scales, the dynamics of the system is equivalent to that of the renormalized normal modes described by Eqs.~\eqref{Qkmf2} and~\eqref{Omk}, with coefficient $\gamma(\varepsilon)$ as in Eq.~\eqref{gamma}, } is the main result of our paper, as announced in the Introduction.
Figure~\ref{fig:gamma} shows a comparison between the {analytical} estimate of $\gamma(\varepsilon)$ {as computed from Eq.~\eqref{gamma}} and its {numerical estimate $\gamma_{num}(\varepsilon)$} in a chain of $N=100$ particles. {The numerical estimate is defined as} {
\begin{equation}
\label{gammaemp}
\gamma_{num}(\varepsilon)=2 g \langle U_2\rangle_{\mu}\overline{\left( \frac{U_4}{U_2^2}\right)}\,,
\end{equation}
where the overbar denotes the empirical time average. The expression {in Eq.~\eqref{gammaemp}} is obtained by plugging the relation $\alpha(\varepsilon)=\frac{N \gamma(\varepsilon)}{2 \langle U_2\rangle_{\mu}}$ (see Eq.~\eqref{gamma}) in the expression of $h_N(\mathbf{q})$ {provided by Eq.~\eqref{hN} and then determining the value of $\gamma(\varepsilon)$ which makes $h_N(\mathbf{q})$ zero on average.}
}

In comparing formulas \eqref{Omk} and \eqref{gamma} with the formula $(11)$ in~\cite{L98}, obtained by another method, one has to take into account the Clausius virial theorem, which yields the asymptotic (for large $N$) identity
\begin{equation}
\label{virial}
g\frac{\langle r^4\rangle_\phi}{\langle r^2\rangle_\phi}\sim \frac{1}{\beta}\frac{1}{\langle r^2\rangle_\phi}-1\ ,
\end{equation}
relating our asymptotic expression for $\gamma$ (left hand side) to that of Lepri (right hand side). 
The asymptotic relation \eqref{virial} is proven in~\ref{Avirial}, where also the link with the quoted works of Casartelli and co-workers emerges. 

\subsection{Heuristic justification of the results}

We conclude this section by giving the heuristic justification of the results presented above. For a rigorous, complete proof the reader is deferred to~\ref{Astat}. 

The quantity $h_N$ \eqref{hN}, written explicitly, reads
\begin{equation}
\label{hNexp}
h_N=g\frac{U_4({\bf q})}{N}-\alpha\frac{[U_2({\bf q})]^2}{N^2}=\frac{1}{N}\sum_{n=0}^{N-1}
\left[gr_n^4-\alpha\left(\frac{1}{N}\sum_{m=0}^{N-1}r_m^2\right)r_n^2\right]\ ,
\end{equation}
where the ``spring variables''
\begin{equation}
\label{rn}
r_n=q_{n+1}-q_n
\end{equation}
are defined for $n=0,\dots,N-1$, together with the fixed ends condition $q_0=q_N=0$. 

As a first step, we observe that the quantity $h_N$ \eqref{hNexp} is a function of the $r_n$ only, so that it seems natural to switch from an integration over the $q_n$ to that on the $r_n$ in the conditioned Gibbs measure \eqref{condGibbs}. {If one neglects the fixed boundary conditions, the change of variables from $\{q_n\}$ to $\{r_n\}$ leads to independent, identically distributed (iid) random variables with the following probability distribution}:
\begin{equation}
\label{rniid}
e^{-\beta(U_2+gU_4)}=e^{-\beta\sum_{n=0}^{N-1}(r_n2/2+gr_n^4/4)}
=\prod_{n=0}^{N-1}e^{-\beta(r_n^2/2+gr_n^4/4)}\ .
\end{equation}
On the other hand, {if the} fixed ends condition {is
  taken into account,} the number of moving $q_n$'s is $N-1$, while
the number of moving $r_n$'s is $N$:  {due to the global constraint}
\begin{equation}
  \sum_{n=0}^{N-1} r_n = q_N-q_0 = 0.
  \label{eq:constraint-q}
\end{equation}
 the $r_n$'s are therefore \textit{not} independent.
{Still, the constraint in Eq.~\eqref{eq:constraint-q} can be
  regarded as a subleading correction, negligible in the large-$N$
  limit, where the $\lbrace r_n \rbrace$ can be for all practical
  purposes treated as iid random variables with probability density
  proportional to the one in Eq.~\eqref{rniid}. In what follows, we will
  therefore work under the hypothesis that the constraint in
  Eq.~\eqref{eq:constraint-q} is negligible and we denote by $\de m$ the {following} Gibbs measure:
\begin{equation}
  \textrm{d}m = \textrm{d}m(r_0,\ldots,r_{N-1}) = \frac{1}{\mathcal{Z}} \prod_{n=0}^{N-1}e^{-\beta(r_n^2/2+gr_n^4/4)}~dr_0\ldots dr_{N-1},
  \label{eq:dm-def}
\end{equation}
where $\mathcal{Z}$ is a normalization factor, and by
$\langle\cdot\rangle_m=\int\cdot\ \de m$ the associated expectation
values.} As a further step, we substitute the arithmetic mean of the
$r_n^2$ inside the round brackets of \eqref{hNexp} with its expectation
with respect to \eqref{rniid}:
\begin{equation}
\label{sumrn2}
{\frac{1}{N}\sum_{n=0}^{N-1}r_n^2\approx  \left\langle \frac{1}{N}\sum_{n=0}^{N-1} r_n^2 \right\rangle_m =
\frac{\int r^2 e^{-\beta(r^2/2+gr^4/4)}\de r}{\int e^{-\beta(r^2/2+g^4/4)}\de r}}=\langle r^2 \rangle_\phi \ .
\end{equation}
{Let us notice that the replacement of the empirical mean over
  the $N-1$ istances of variables $r_n$ with the average over the
  equilibrium distribution of a single $r_n$ written in
  Eq.~\eqref{sumrn2} is exact in the large-$N$ limit, due to the strong
  law of large numbers, as proved in~\ref{Astat}. By
  exploiting the above approximations we can then write the normalized
  difference $h_N = (H -\mathcal{H})/N$ between our original and mean
  field Hamltonian in term of the $r_n$ variables as}
\begin{equation}
\label{hNapp}
h_N(r_0,\ldots,r_{N-1})\approx \frac{1}{N}\sum_{n=0}^{N-1}
\left(gr_n^4-\alpha\langle r^2\rangle_\phi r_n^2\right)\equiv \frac{1}{N}\sum_{n=0}^{N-1} y_n,
\end{equation}
{where we have introduced a set of new iid random variables
  $y_n$:
\begin{equation}
  y_n = gr_n^4-\alpha\langle r^2\rangle_\phi r_n^2.
  \label{eq:yn-def}
\end{equation}
It can be easily realized that a sufficient condition for
\begin{equation}
  \langle h_N(r_0,\ldots,r_{N-1})\rangle_m = 0,
  \label{eq:meanhn-0}
\end{equation}
is $\langle y_n\rangle = 0$ $\forall n$. This can be obtained in turn by fixing the coefficient $\alpha$
that appears in the definition of $y_n$, Eq.~\eqref{eq:yn-def}, to
\begin{equation}
  \alpha(\beta)=g\langle r^4\rangle_\phi/\langle r^2\rangle_\phi
  \label{eq:alpha-largeN-2}
\end{equation}
The expression obtained for $\alpha(\beta)$ in
Eq.~\eqref{eq:alpha-largeN-2} is the
large-$N$ limit of $\alpha(\beta)$ written in Eq.~\eqref{alphaeps},
namely the one associated to the strong convergence condition for
$h_N$ of Eq.~\eqref{mustrong}. With the choice of $\alpha(\beta)$ in
Eq.~\eqref{eq:alpha-largeN-2} we have now that the normalized energy
difference $h_N(r_0,\ldots,r_{N-1})$ is the arithmetic mean of the
$N$, zero mean, iid random variables $y_n$. All the moments $\langle |y_n|^k\rangle_\phi$ exist,
for any $k\geq0$, and in particular for $k=2,4$. Let us denote by $m\left(\lbrace|h_N|>\eta\rbrace\right)$, where $\eta>0$
is a positive real number, the probability to sample a value of $h_N$
in absolute value larger than $\eta$ according to the measure
$\textrm{d}m$, namely
\begin{equation}
  m\left(\left\{|h_N|>\eta\right\}\right) =
  \int \textrm{d}m(r_0,\ldots,r_{N-1})~\Theta(|h_N(r_0,\ldots,r_{N-1})|-\eta).
\end{equation}
Due to the Chebyshev inequality, one gets
\begin{equation}
\label{heumuweak}
m\left(\left\{|h_N|>\eta\right\}\right)\leq\frac{\langle h_N^2\rangle_m}{\eta^2}=
\frac{\langle y_n^2\rangle_\phi}{N\eta^2}\, .
\end{equation}
The bound in Eq.~\eqref{heumuweak} on the probability distribution of
$|h_N|$ implies the weak convergence result of
Eq.~\eqref{muweak}, as soon as one sets $\eta=N^{-a}$ with $0< a<1/4$.
Under the same hypothesis, just exploited, that the $r_n$'s are iid
random variables, it is
possible to derive an even stronger bound on the probability
distribution of $|h_N|$ by applying the Markov inequality
\cite{Ross}:
\begin{equation}
\label{heumustrong}
m\left(\left\{|h_N|>\eta\right\}\right)\leq\frac{\langle h_N^4\rangle_m}{\eta^4}=
\frac{N\langle y_n^4 \rangle_\phi+3N(N-1)\langle y_n^2\rangle_\phi^2}{N^4\eta^4}\, .
\end{equation}
The latter upper bound goes to zero as $1/N^2$ for any $\eta>0$, which
is enough \cite{Ross} to imply a strong convergence law
$m(\{\lim_{N\to\infty}h_N=0\})=1$. This concludes our {\it
  heuristic} derivation of the bounds written in Eq.~\eqref{muweak} and
Eq.~\eqref{mustrong}.\\ \\}

{As an addendum to the heuristic derivation of both the weak
  and strong convergence results for the difference $h_N$ between the
  original Hamiltonian of the systems, $H$ in Eq.~\eqref{H}, and the
  mean-field one proposed in this paper, $\mathcal{H}$ in
  Eq.~\eqref{Hmf}, let us also sketch the argument to heuristically
  prove Eq.~\eqref{U2law} and Eq.~\eqref{U2lim}, which are crucial to
  obtain the renormalized frequency spectrum of the effective normal
  modes. To this purpose, let us go back to the substitution of the
  arithmetic mean $\sum_{n=0}^{N-1} r_n^2/N$ with the expectation over
  the probability distribution of the single variable written in
  Eq.~\eqref{sumrn2}. If we then recall that
  $\sum_{n=0}^{N-1}r_n^2/N=2U_2/N$, demostrating the relation in Eq.~\eqref{sumrn2} is equivalent to proving the
  relations in Eqs.~\eqref{U2law} and~\eqref{U2lim}. The relation
  in Eq.~\eqref{U2lim} is readily obtained by noticing that
\begin{equation}
\left\langle \frac{U_2}{N}\right\rangle_m=\frac{1}{2}\langle r^2\rangle_\phi.
\end{equation}
For what concerns then the relation in Eq.~\eqref{U2law}, let us notice
that we can write the energy difference per degree of freedom $(U_2-\langle
U_2\rangle_m)/N$ as the arithmetic mean of the iid, zero mean random variables
\begin{equation}
\frac{U_2-\langle U_2\rangle_m}{N}=
\frac{1}{N}\sum_{n=0}^{N-1}\frac{r_n^2-\langle r^2\rangle_\phi}{2}\equiv
\frac{1}{N}\sum_{n=0}^{N-1}z_n\ ,
\end{equation} 
where $z_n=(r_n^2-\langle r^2\rangle_\phi)/2$. Repeating the arguments
of the above paragraphs for $h_N$, one obtains in a similar way a heuristic derivation of the expression in Eq.~\eqref{U2law}.}

\section{Stochastic dynamics of the effective normal modes}
\label{sec-3}




At this stage we have shown that the dynamics of the system is close, in the large $N$ limit and in a statistical sense, to that of a mean field model that can be reduced to $N-1$ independent, effective normal modes. The energies of the latter modes are not expected to be conserved along the flow of the original system.
An effective, phenomenological representation of the true
microscopic dynamics consistent with the given picture is the one already suggested in \cite{L98}, namely
\begin{equation}
  \ddot{Q}_k(t)= -\Omega_k^2 Q_k(t) -\eta_k \dot{Q}_k(t) + 
  \sqrt{2\varepsilon\eta_k}\ \xi_k(t)\ ,
  \label{Langevin}
\end{equation}
where $\Omega_k=\sqrt{1+\gamma(\varepsilon)}\ \omega_k$, $\gamma(\varepsilon)$ given in \eqref{gamma}, $\xi_k(t)$ is the standard, delta correlated Gaussian process satisfying
\begin{equation}
\langle\xi_k(t)\rangle=0\ ;\  \langle \xi_k(t) \xi_q(s)\rangle = \delta_{kq}~\delta(t-s)\ ,
  \label{xiprocess}
\end{equation}
$\langle\cdot\rangle$ denoting, here and henceforth, average over the stochastic process. In Eq.~\eqref{Langevin} one assumes that $ \langle Q_k^2\rangle \Omega_k^2 \simeq \varepsilon$: numerical computations show that this relation is approximately valid also in a non-perturbative regime, for $\varepsilon \simeq O(1)$. The problem of how to choose the damping coefficient $\eta_k$ appearing in \eqref{Langevin} is nontrivial. We assume the form
\begin{equation}
\label{etak}
\eta_k=c(\varepsilon)\Omega_k^{5/3}=c(\varepsilon)(1+\gamma(\varepsilon))^{5/6}\omega_k^{5/3}\ .
\end{equation}
The latter expression is a tentative generalization of the one deduced in \cite{L98} in the hydrodynamic limit,
i.e. as $k/N\to0$.  We will come back on the coefficient $c(\varepsilon)$ below. 
As is well known \cite{Chandra}, the stochastic oscillator equation \eqref{Langevin} is such that the expected value of the $k$-th effective harmonic energy \eqref{Ekeff} converges exponentially fast to $\varepsilon$, namely
\begin{equation}
\langle\mathcal{E}_k(t)\rangle\to\varepsilon
\end{equation}
as $t\to\infty$, on a time-scale $1/\eta_k$, for any $k=1,\dots,N-1$ and any initial condition. 

One can try to deduce the Langevin equation \eqref{Langevin} from first principle starting
from the original microscopic Hamiltonian dynamics and applying the
Zwanzig-Mori projection formalism~\cite{L98} or by using a
wave-turbulence approximation scheme~\cite{GLC05,GLC07,K49}. Quite
unfortunately, in both cases some {\it bone-fide} approximations must
be made on the closure of a hierarchy of multipoint correlation
functions in order to end up with a Langevin effective dynamics as
simple as \eqref{Langevin}.  Although the whole
procedure is well motivated and discussed in both cases, the final
derivation is somehow heuristic. According to the present state of the art, the Langevin equations \eqref{Langevin} for the effective normal modes must be assumed simply as phenomenologically reasonable guess to be compared 
with the numerical data obtained from the original Hamiltonian dynamics, neamely
\begin{equation}
  \ddot{q}_n = - \frac{\partial}{\partial q_n}\left[ U_2({\bf q}) + gU_4({\bf q})\right]
  \label{Hameqns}
\end{equation}
\begin{figure}
    \centering
    \includegraphics[width=.69\linewidth]{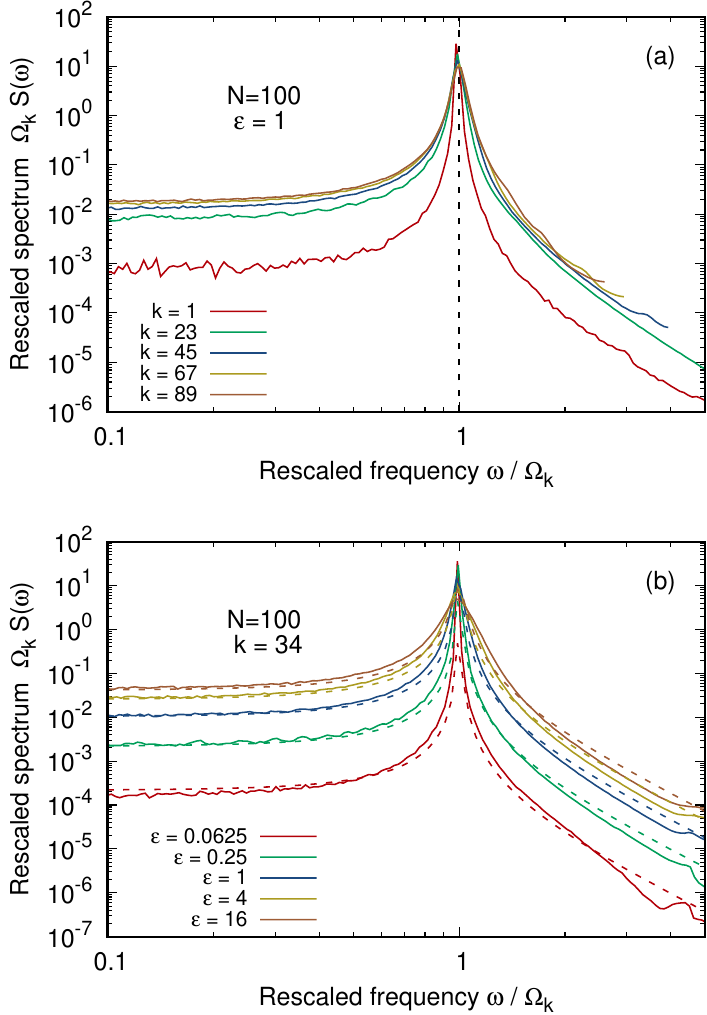}
    \caption{Power density spectra of the normal modes $Q_k$'s. In panel (a) different values of $k$ are considered, at specific energy $\varepsilon=1$. Frequencies are rescaled by $\Omega_k$: as expected, the peaks of the spectra are found in 1. Panel (b) shows instead the behaviour of the spectrum for $k=34$ at different values of the energy. The dashed lines are obtained by fitting the functional form~\eqref{SkomL2} (with $c_0$ set to the phenomenological value $0.4$ and one free parameter for normalization). Details on the numerical simulations as in Fig.~\ref{fig:gamma}.}
    \label{fig:spectra}
\end{figure}
To such a purpose we have simulated the dynamics generated
by the original Hamilton equations \eqref{Hameqns}, measuring the power spectrum
\begin{equation}
\label{Skom}
S_k(\omega) = 
\lim_{T\to+\infty}\frac{1}{T}\left\langle\left| \int_0^T e^{- i \omega t}Q_k(t)\ \de t\right|^2\right\rangle
\end{equation}
of each actual Fourier mode $Q_k(t)=\sqrt{2/N}\sum_{n=1}^{N-1}q_n(t)\sin(\pi kn/N)$. As is well known \cite{Chandra,WU}, if $Q_k(t)$ solves equation \eqref{Langevin}, its (normalized) power spectrum is exactly given by
\begin{equation}
\label{SkomL}
S_k(\omega)=\frac{G_k(\omega)}{\int_0^{\infty} d \omega' \, G_k(\omega')}\ ,
\end{equation}
with
\begin{equation}
\label{SkomL2}
G_k(\omega)=\frac{2\eta_k\varepsilon}{(\omega^2-\Omega_k^2)^2+\eta_k^2\omega^2}\ .
\end{equation}
This theoretical power spectrum is compared with the numerical one in Fig.~\ref{fig:spectra}. 

\begin{figure}
    \centering
    \includegraphics[width=.7\linewidth]{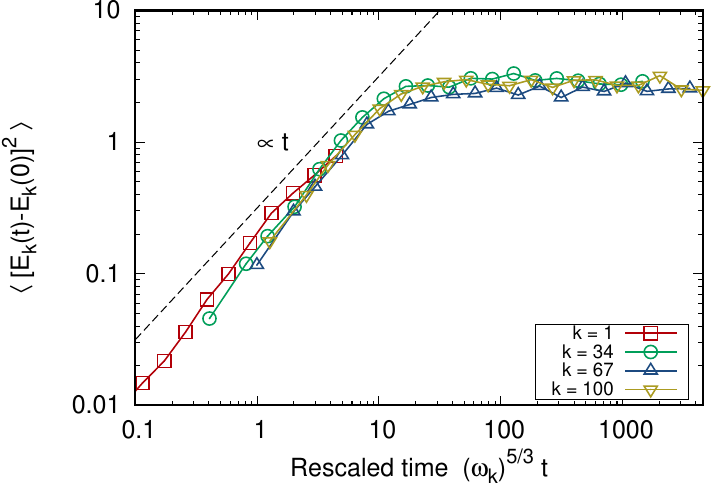}
    \caption{Mean square deviation of the spectral energies, for different values of $k$. Each curve has been obtained by averaging over 500 branches of a single trajectory of the system. Each branch spans a time window of 8 periods of the slowest normal mode. Here $\varepsilon=1$. The dashed line, proportional to $t$, is a guide for the eyes. Further details on the numerical simulations as in Fig.~\ref{fig:gamma}.}
    \label{fig:msd}
\end{figure}
Concerning the multiplicative factor $c(\varepsilon)$ of the damping coefficient \eqref{etak}, we took
\begin{equation}
\label{ceps}
c(\varepsilon)=c_0\ \frac{(g\varepsilon)^{4/3}}{[1+\gamma(\varepsilon)]^3}\ ,
\end{equation}
where $c_0 \simeq 0.4$ is a numerical constant (determined heuristically). Such an expression was obtained by extending to the whole frequency spectrum what found by Lepri in the hydrodynamic limit \cite{L98}. 

The rather good agreement of the theoretical and numerical power spectra of each Fourier mode, as shown in Fig.~\ref{fig:spectra}, is a clear signature of the fact that the Langevin equations \eqref{Langevin} describe the actual dynamics of the system, notwithstanding the complex network of
interactions between the actual modes. The stochastic oscillator equations \eqref{Langevin} are linear and independent of each other, which definitely justifies the following conclusion: the dynamics of the system is decomposed, to all energy scales, into $N-1$ effective normal modes.    

As a further numerical test, in Fig.~\ref{fig:msd} we plot the mean square deviation of the energy of the 
$k$-th Fourier mode versus $\eta_kt$, for any $k$. The ``collapse'' of the curves, predicted by the equations \eqref{Langevin}, is another signature of the effective normal modes.     

\section{Conclusions}
\label{sec-4}

{Let us summarize in this final section the main results of the
  paper. First, we have proposed a {\it mean-field} Hamiltonian
  $\mathcal{H}$ for FPUT models with quartic interactions, of the form
  $\mathcal{H} = H + N h_{N}$, where $H$ is the original Hamiltonian,
  $N$ is the number of degrees of freedom in the system and $h_N$ is a
  reminder which is negligible in measure in the thermodynamic
  limit. The idea is to have {\it typical} equilibrium properties,
  according to the canonical ensemble measure, which are very close
  for $H$ and $\mathcal{H}$. The technical part of our work consisted
  in proving the convergence $h_N \rightarrow 0$ for
  $N\rightarrow\infty$, both in the weak and in the strong
  probabilistic sense. The choice of the mean-field Hamiltonian
  $\mathcal{H}$ in Eq.~\eqref{Hmf} explicitly shows that in the
  thermodynamic limit the dynamics of the FPUT model with quartic
  interactions is close to the dynamics of $N$ independent harmonic
  oscillators with renormalized frequencies
  \begin{equation}
    \Omega_k = \sqrt{1+\gamma(\varepsilon)}~\omega_k\,,
  \end{equation}
  where $\omega_k$ are the original frequencies of the Fourier modes.
  The coefficient $\gamma(\varepsilon)$ depends on the energy per degree of freedom,
  $\varepsilon = E/N$, and can be analytically computed (see for
  instance the comparison between the analytical prediction and the
  numerical estimate for $\gamma(\varepsilon)$ in Fig.~\ref{fig:gamma}
  of Sec.~\ref{sec:results-implications}). We have then validated the analytical proof that effective normal
  modes with renormalized frequencies are a good description of the
  system at all energies by comparing the numerical simulations of the
  original non-linear FPUT chain with the dynamics of uncoupled
  effective stochastic equations, Eq.~\eqref{Langevin}, built from the
  renormalized normal modes picture.}

  {Overall, the scenario drawn by our results is somehow
    different from  the standard literature on FPUT models,
    which mostly focuses on the slow relaxation to
    equilibrium for choices of the model parameters perturbatively close  to the integrable limits. We have pursued, on the
    contrary, a large-$N$, fully non-perturbative approach, showing that
    {\it at all energy scales} the equilibrium distribution of the
    model and its dynamics are well approximated by that of
    independent harmonic oscillators with renormalized
    frequencies. In doing so, we have shown that the probabilistic approach to the
    model phenomenology is a possible path to reveal the emergence
    of a sort of  ``effective integrability'' of the model at all
    energy scales. Our results also corroborate the viewpoint
    according to which chaos has a marginal role for the emergence of
    statistical properties, while the large-$N$ limit is crucial (see also the recent review paper~\cite{BGV21} by some of the authors). Let us consider for instance
    Fig.~\ref{fig:msd} in Sec.~\ref{sec-3}, where it is shown that the
    equipartition time for the $k$-th effective mode with renormalized
    frequency $\Omega_k$ is proportional to the inverse of the damping
    coefficient $\eta_k$ entering the effective dynamics in
    Eq.~\eqref{Langevin}. The timescale
    $\tau_k \sim \eta_k^{-1}$, in particular for small $k$, is much
    larger than the Lyapunov time: this represents a strong hint that for the system, also in the strongly chaotic
    regimes, chaos has a marginal role for the dynamics of the
    large-scale modes. As a perspective it would not be difficult to
    prove that our main result, namely the existence of an effective
    Hamiltonian close, in a statistical sense, to the true one,
    extends to the case where $U_2+U_4$ is replaced with a generic
    $\sum_nF(q_n-q_{n-1})$ where $F(r)$ is an even function of its
    argument, but we leave this for a future work.\\ As a side
    observation, it is finally interesting to note that the
    equivalence ``in measure'' of the FPUT models to an integrable
    model, bears a remarkable similarity with a property known as {\it
      triviality} in quantum/statistical field theories, according to
    which non-linear/non-integrable theory turns out to be fully
    equivalent to a free theory (integrable by definition) upon
    summing up over all quantum/statistical fluctuations. This is for
    instance what happens in certain physical dimensions for a scalar
    field theory with quartic self-interaction, see, e.g.,
    ~\cite{PS95} or~\cite{P88,L91}.}

\section*{Acknowledgements}
MB was supported by ERC Advanced Grant RG.BIO (Contract No. 785932).
{GG acknowledges partial support by the project MIUR-PRIN2022,
``Emergent Dynamical Patterns of Disordered Systems with
Applications to Natural Communities'', code 2022WPHMXK.}

\appendix

\section{Proofs of the main results}
\setcounter{section}{1}
\label{Astat}

\subsection{Conditioned canonical averages}

The statistical mechanics of the FPUT systems with fixed ends is defined by the canonical Gibbs measure $\de\mu=\frac{1}{Z}e^{-\beta H(q,p)}|_{q_0=q_N=0}\de q_1\dots\de q_{N-1}\de p_1\dots\de p_{N-1}$, introduced in \eqref{condGibbs}. We need to compute only the expectations of functions that, like the Hamiltonian $H$ of the system, depend on the canonical coordinates only through the relative displacements, or spring variables $r_n=q_{n+1}-q_n$, $n=0,\dots,N-1$. Then, it is clearly convenient to make use of the $r_n$ as integration variables, which has to be done taking into account the geometric constraint 
\begin{equation}
\sum_{n=0}^{N-1}r_n=\sum_{n=0}^{N-1}(q_{n+1}-q_n)=q_N-q_0=0\ .
\end{equation}
Such a conditioning is obtained, when averaging, by setting $q_0=0$ (so that $r_0=q_1$, $r_1=q_2-q_1$ and so on), and integrating 
$e^{-\beta H}$ against 
\begin{eqnarray}
&&\delta(q_N)\de q_1\dots\de q_N=\delta(q_N)
\det\frac{\partial(q_1,\dots,q_N)}{\partial(r_0,\dots,r_{N-1})}\de r_0\dots\de r_{N-1}
=\nonumber\\
&&=\delta\left(\sum_{n=0}^{N-1}r_n\right)\de r_0\dots\de r_{N-1}\ .
\label{jacob}
\end{eqnarray}
This leads to the following conclusion: the average $\langle f\rangle_\mu$ of any function $f(r,p)=f(r_0,\dots,r_{N-1},p_1,\dots,p_{N-1})$ depending, like the Hamiltonian
$H$, on the coordinates only through the spring variables, is given by
\begin{equation}
\label{avfr}
\langle f\rangle_\mu=\frac{1}{Z}\int f(r,p)
e^{-\beta H(r,p)}\delta\left(\sum_{n=0}^{N-1}r_n\right)
\de r_0\dots\de r_{N-1}\de p_1\dots\de p_{N-1}\ ,
\end{equation}
where the partition function $Z$ is such that $\langle 1\rangle_\mu=1$. Now, making use of the explicit expression of the Hamiltonian, i.e.
\begin{equation}
\label{Hsum}
H=\sum_{n=1}^{N-1}\frac{p_n^2}{2}+\sum_{n=0}^{N-1}\phi(r_n)\ ;\ \ 
\phi(r)=\frac{r^2}{2}+g\frac{r^4}{4}\ ,
\end{equation}
and making use of the integral representation  $\delta(x)=\frac{1}{2\pi}\int 
e^{\imath\lambda x}\de\lambda$ of the delta function, one gets
\begin{equation}
\label{avfr2}
\langle f\rangle_\mu=\frac{1}{2\pi Z}\int f(r,p)\prod_{j=1}^{N-1}e^{-\beta\frac{p_j^2}{2}}\de p_j
\prod_{n=0}^{N-1}e^{-\beta\phi(r_n)+\imath\lambda r_n}\de r_n\de\lambda
\end{equation}
with the following explicit expression of the partition function:
\begin{eqnarray}
\label{Zexp}
Z&=&\frac{1}{2\pi}\int \prod_{j=1}^{N-1}e^{-\beta\frac{p_j^2}{2}}\de p_j
\prod_{n=0}^{N-1}e^{-\beta\phi(r_n)+\imath\lambda r_n}\de r_n\de\lambda=\nonumber\\
&=&\left(\frac{2\pi}{\beta}\right)^{\frac{N-1}{2}}\zeta_\phi^N(\beta)
\int\langle e^{\imath\lambda r} \rangle_\phi^N\ \frac{\de\lambda}{2\pi}\ .
\label{Z}
\end{eqnarray}
In the latter expression, we have introduced the quantities
\begin{equation}
\label{springav}
\zeta_\phi(\beta):=\int e^{-\beta \phi(r)}\de r\ \ ;\ \ 
\langle\cdot\rangle_\phi:=\frac{\int (\cdot) e^{-\beta\phi(r)}\de r}{\int e^{-\beta\phi(r)\de r}}\ ,
\end{equation}
namely, the partition function and the average relative to the single spring, respectively.
Formulas \eqref{avfr2}-\eqref{springav} are the key ingredients to compute expectations of functions of spring variables and momenta in the FPUT problem with fixed ends.  

\subsection{Spring correlations}

In the sequel we will need to compute averages of products of powers of (a finite number of) spring variables, and to study their limit as $N\to\infty$. We thus start by focussing on the correlation coefficients $\langle r_0^{a_0}\cdots r_{N-1}^{a_{N-1}}\rangle_\mu$, 
where $a_1,\dots,a_N$ is any sequence of integer nonnegative exponents. 
By exploiting formula \eqref{avfr2}, one easily gets
\begin{equation}
\label{Cexp}
\langle r_0^{a_0}\cdots r_{N-1}^{a_{N-1}}\rangle_\mu=\frac{\int\prod_{j=0}^{N-1}\langle r^{a_j}e^{\imath\lambda r}\rangle_\phi\de\lambda}{\int\langle e^{\imath\lambda r}\rangle_\phi^N \de\lambda}\ .
\end{equation}

\noindent
We now study the asymptotic behaviour of \eqref{Cexp} as $N\to\infty$ restricting to the case of a {\it finite number $p$ of even exponents different from zero}. More precisely, we suppose that $a_j\neq0$ is even if $j=0,\dots,p-1$, while $a_j=0$ if $j=p,\dots,N-1$, where $p\geq1$ is a fixed number independent of $N$. This is the case we need in the present paper. Of course, more general estimates may be obtained as well. With the hypotheses made, formula \eqref{Cexp} gives
\begin{equation}
\label{Cp}
\langle r_0^{a_0}\dots r_{p-1}^{a_{p-1}}\rangle_\mu
=\frac{\int\prod_{j=0}^{p-1}\langle r^{a_j}\cos(\lambda r)\rangle_\phi
\langle\cos(\lambda r)\rangle_\phi^{N-p}
\de\lambda}{\int\langle \cos(\lambda r)\rangle_\phi^N \de\lambda}\ .
\end{equation}
In both the numerator and the denominator of the above formula, the only quantity containing $N$ is $\langle \cos(\lambda r)\rangle_\phi^N$, an even function of $\lambda$ displaying an absolute maximum of unit value at $\lambda=0$ and decaying faster than any power of $\lambda$ as $\lambda\to\infty$. Setting 
$\lambda=\xi/\sqrt{\langle r^2\rangle_\phi N}$ and expanding the cosine, one gets
\begin{eqnarray}
\left\langle \cos\left(\frac{\xi r}{\sqrt{\langle r^2\rangle_\phi N}}\right)\right\rangle_\phi^N&=&
\left(1-\frac{\xi^2}{2N}+
\frac{\xi^4\langle r\rangle_\phi^4}{24N^2\langle r^2\rangle_\phi^2}+\cdots\right)^N
=\nonumber\\
&=& e^{N\ln\left(1-\frac{\xi^2}{2N}+
\frac{\xi^4\langle r\rangle_\phi^4}{24N^2\langle r^2\rangle_\phi^2}+\cdots\right)}=
\nonumber\\
&=&e^{-\frac{\xi^2}{2}+\frac{\xi^4}{24N}
\frac{\langle r^4\rangle_\phi-3\langle r^2\rangle_\phi^2}{\langle r^2\rangle_\phi^2}+\cdots}=
\nonumber\\
&=& e^{-\frac{\xi^2}{2}}\left(1+\frac{\xi^4}{24N}
\frac{\langle r^4\rangle_\phi-3\langle r^2\rangle_\phi^2}{\langle r^2\rangle_\phi^2}+\cdots\right)
\end{eqnarray}
as $N\to\infty$, where dots denote $O(1/N^2)$ terms. Now, making the same change of variable $\lambda=
\xi/\sqrt{\langle r^2\rangle_\phi N}$ in the integral \eqref{Cp}, and carefully retaining terms of order $1/N$ everywhere, after a long elementary calculation similar to the one reported above, yields
\begin{equation}
\langle r_0^{a_0}\dots r_{p-1}^{a_{p-1}}\rangle_\mu
=\prod_{j=0}^{p-1}\langle r^{a_j}\rangle_\phi\left[
1+\frac{p}{2N}\left(1-\frac{1}{p}\sum_{j=0}^{p-1}
\frac{\langle r^{a_j+2}\rangle_\phi}{\langle r^2\rangle_\phi\langle r^{a_j}\rangle_\phi}\right)
+\cdots\right]\ , \label{Cpas}
\end{equation}
showing that $\langle r_0^{a_0}\dots r_{p-1}^{a_{p-1}}\rangle_\mu-\prod_{j=0}^{p-1}\langle r^{a_j}\rangle_\phi=O(1/N)$ as $N\to\infty$, i.e. the spring variables are almost uncorrelated in the thermodynamic limit, as expected. The sub-leading $O(1/N)$ term in \eqref{Cpas} is retained in order to show, first, that the hypothesis of finite $p$, i.e. $p$ independent of $N$, is crucial to our purposes, for otherwise a finite fraction of spring variables would turn out to have a finite correlation, and, second, because we will need its explicit expression below.

\subsection{Asymptotic estimates}

We start by computing the canonical averages of the relevant physical quantities in the thermodynamic limit ($N\to\infty$). We will make use of the general formulas \eqref{avfr2}--\eqref{Zexp} and of the correlation formula \eqref{Cpas}. 
To later purposes, it is convenient to define, for any even integer $a=2,4,\dots$, the specific potential energies of degree $a$, namely
\begin{equation}
\label{ua}
u_a\equiv \frac{1}{aN}\sum_{n=0}^{N-1}r_n^a\ .
\end{equation}
Thus $u_2=U_2/N$, $u_4=U_4/N$ and so on. By \eqref{Cpas}, as $N\to\infty$, we get
\begin{equation}
\label{uaav}
\langle u_a\rangle_\mu= \frac{\langle r^a\rangle_\mu}{a}\to
\frac{\langle r^a\rangle_\phi}{a}\ ;
\end{equation} 
\begin{equation}
\label{ua2av}
\langle u_a^2 \rangle_\mu=\frac{1}{a^2N^2}\left[N\langle r^{2a}\rangle_\mu+
N(N-1)\langle r_0^ar_1^a \rangle_\mu\right]\to\frac{\langle r^a\rangle_\phi^2}{a^2}\ .
\end{equation}  
By such formulas, observing that, according to \eqref{avfr2}, the momenta are Gaussian distributed with zero mean and variance $1/\beta$, we obtain the
specific energy of the system, namely 
\begin{eqnarray}
\varepsilon(\beta)&=&\frac{\langle H \rangle_\mu}{N}=
\frac{N-1}{N}\frac{1}{2\beta}+\langle u_2\rangle_\mu+g\langle u_4\rangle_\mu=
\nonumber\\
&=& \frac{N-1}{N}\frac{1}{2\beta} +\frac{\langle r^2\rangle_\mu}{2}+
g\frac{\langle r^4\rangle_\mu}{4}\to\frac{1}{2\beta}+\frac{\langle r^2\rangle_\phi}{2}+
g\frac{\langle r^4\rangle_\phi}{4}=\nonumber\\
&=&\frac{1}{2\beta}+\langle \phi(r) \rangle_\phi \label{epslim}
\end{eqnarray}
as $N\to \infty$. Moreover, the large $N$ limit of the coefficient $\alpha(\beta)$ defined in 
\eqref{alphabeta} is
\begin{equation}
\label{alphalim}
\alpha(\beta)=g\frac{N\langle U_4\rangle_\mu}{\langle U_2^2\rangle_\mu}=
g\frac{\langle u_4\rangle_\mu}{\langle u_2^2\rangle_\mu}\to 
g\frac{\langle r^4\rangle_\phi}{\langle r^2\rangle_\phi^2}\ ,
\end{equation}
where \eqref{uaav} and \eqref{ua2av} have been used; \eqref{epslim} and 
\eqref{alphalim} coincide with \eqref{alphaeps}. 

Due to formulas \eqref{Cpas}, \eqref{uaav} and \eqref{ua2av}, the asymptotic value of the second moment of $u_a$ turns out to be
\begin{eqnarray}
\langle (u_a-\langle u_a \rangle_\mu)^2\rangle_\mu&=&
\frac{\langle r^{2a}\rangle_\mu-\langle r_0^ar_1^a\rangle_\mu+
N(\langle r_0^ar_1^a\rangle_\mu-\langle r^a\rangle_\mu^2)}{Na^2}=\nonumber\\
&=&\frac{\langle r^{2a}\rangle_\phi\langle r^2\rangle_\phi-
\langle r^a\rangle_\phi\langle r^{a+2}\rangle_\phi}{a^2N\langle r^2\rangle_\phi}+
O(1/N^2)\ . \label{ua2mom}
\end{eqnarray}
In particular, we observe that for $a=2$ the formula above implies that
$\langle (u_a-\langle u_a \rangle_\mu)^2\rangle_\mu=O(1/N^2)$, which, by the Chebyshev inequality \cite{JP}, gives 
\begin{equation}
\mu\left(\left\{({\bf q},{\bf p}): |u_2({\bf q})-\langle u_2\rangle_\mu|>\eta\right\}\right)=O(1/N^2)\ .
\end{equation}
The latter measure is summable in $N$, which implies \cite{Gned} the almost sure, or strong convergence result \eqref{U2law}.

We can now prove the main result of the paper, namely the convergence to zero of 
the Hamiltonian perturbation $h_N=g u_4-\alpha u_2^2$, namely \eqref{muweak} and \eqref{mustrong}. In order to prove \eqref{muweak}, we start by observing that 
$\langle h_N\rangle_\mu=0$. Thus, by the Chebyshev inequality:
\begin{eqnarray}
\mu(\{|h_N|>\eta\})&\leq&
\frac{\langle (h_N-\langle h_N\rangle_\mu)^2\rangle_\mu}{\eta^2}
=\nonumber\\
&=&\frac{\langle [g(u_4-\langle u_4\rangle_\mu)-
\alpha (u_2^2-\langle u_2^2\rangle_\mu)]^2\rangle_\mu}{\eta^2}\leq\nonumber\\
&\leq&2\left[\frac{g^2\langle (u_4 -\langle u_4\rangle_\mu)^2\rangle_\mu}{\eta^2}+
\frac{\alpha^2\langle (u_2^2-\langle u_2^2\rangle_\mu)^2\rangle_\mu}{\eta^2}\right]\ ,
\label{hNweak}
\end{eqnarray}
where we used $2\langle AB\rangle_\mu\leq
\langle A^2\rangle_\mu+\langle B^2\rangle_\mu$, true for any pair of functions $A$ and $B$, since $\langle (A-B)^2\rangle_\mu\geq 0$. Now, the first term in square brackets on the last line of \eqref{hNweak} is $O(1/N)$ by
\eqref{ua2mom}. The second term is also $O(1/N)$, since
\begin{eqnarray}
\langle (u_2^2-\langle u_2^2\rangle_\mu)^2\rangle_\mu&=&\langle u_2^4\rangle_\mu-
\langle u_2^2\rangle_\mu^2=\nonumber\\
&=&\frac{\sum_{ijkl}\langle r_i^2r_j^2r_k^2r_l^2\rangle_\mu-
\sum_{ijkl}\langle r_i^2r_j^2\rangle_\mu\langle r_k^2r_l^2\rangle_\mu}{16N^4}\ ,
\end{eqnarray}
and by \eqref{Cpas} the leading term $O(N^4)$ in the numerator vanishes. Thus,
\eqref{hNweak} implies 
$\mu(\{|h_N|>\eta\})=O(1/(\eta^2N))$, which proves \eqref{muweak} by choosing 
$\eta=N^{-a}$, with $0<a<1/2$.

In order to prove \eqref{mustrong}, we first observe that $h_N=gu_4-\alpha u_2^2$ is a linear combination of the random variables $u_4$ and $u_2^2$. Since $u_2$ converges almost everywhere to its average, and the same holds for $u_2^2$, the square being a continuous function \cite{JP}, if also $u_4$ converges almost surely to its average,
then $h_N$ almost surely converges to its average, i.e. zero (this is obtained by observing that the union of the zero measure sets where $u_4$ and $u_2^2$ do not converge has zero measure). In order to prove that $u_4-\langle u_4\rangle_\mu$ converges almost everywhere to zero, we compute its fourth moment. We actually compute the fourth moment of $u_a$, where $a$ is any even positive integer. By collecting the 
various correlation terms appearing in $\langle (\sum_ir_i^a)^s\rangle_\mu$, for $s=1,2,3,4$, which requires a combinatorial counting, after a long calculation, we find
\begin{eqnarray}
\langle (u_a-\langle u_a\rangle_\mu)^4\rangle_\mu&=&
\langle u_a^4\rangle_\mu-4\langle u_a^3\rangle_\mu\langle u_a\rangle_\mu+
6\langle u_a^2\rangle_\mu\langle u_a\rangle_\mu^2-3\langle u_a\rangle_\mu^4=
\nonumber\\
&=& \frac{1}{a^4}\left[ A+\frac{B}{N}+O(1/N^2)\right]\ ,\label{ua4mom}
\end{eqnarray}
where 
\begin{equation}
A=\langle r_0^ar_1^ar_2^ar_3^a\rangle_\mu -4\langle r^a\rangle_\mu\langle r_0^ar_1^ar_2^a\rangle_\mu+6\langle r^a\rangle_\mu^2\langle r_0^ar_1^a\rangle_\mu
-3\langle r^a\rangle_\mu^4\ ;
\end{equation}
\begin{eqnarray}
B&=&-6\langle r_0^ar_1^ar_2^ar_3^a\rangle_\mu+6\langle r_0^{2a}r_1^ar_2^a\rangle_\mu
+12\langle r^a\rangle_\mu\langle r_0^ar_1^ar_2^a\rangle_\mu+\nonumber\\
&-&12\langle r^a\rangle_\mu\langle r_0^{2a}r_1^a\rangle_\mu
+6\langle r^a\rangle_\mu^2(\langle r^{2a}\rangle_\mu-
\langle r_0^ar_1^a\rangle_\mu)\ .
\end{eqnarray}
By using \eqref{Cpas} to leading order, one easily checks that $B=O(1/N)$. On the other hand, taking into account the $O(1/N)$ correction in \eqref{Cpas}, one finds that
$A=O(1/N^2)$. Thus, \eqref{ua4mom} and the Markov inequality imply
\begin{equation}
\mu(\{|u_a-\langle u_a\rangle_\mu|>\eta\})\leq
\frac{\langle (u_a-\langle u_a\rangle_\mu)^4\rangle_\mu}{\eta^4}=O(1/N^2)\ .
\end{equation} 
Since the latter measure is summable in $N$, $u_a-\langle u_a\rangle_\mu$ almost surely converges to zero for any even $a$. This implies, by the above remarks, \eqref{mustrong}.

As a final comment, the almost sure convergence of $u_a$ - which implies the convergence in probability - allows to export our result to any even polynomial potential $\phi(r)$, replacing any potential term $u_a$ by $u_2^{a/2}$ times a suitable coefficients similar to 
$\alpha$ in the mean field Hamiltonian.

{Similar techniques have been used in~\cite{MBC} to control the quasi independence of the spring variables.}

\section{Mean-field Hamiltonian vs. Birkhoff normal form as $\varepsilon\to0$}
\setcounter{section}{2}
\label{Apert}

The low temperature ($\beta\to\infty$) limit of our mean field theory can be explicitly computed. Since, for any integer $k\geq1$, as $\beta\to\infty$
\begin{eqnarray}
\langle r^{2k}\rangle_\phi&=&\frac{\int r^{2k} e^{-\beta(r^2/2+gr^4/4)}\de r
}{\int e^{-\beta(r^2/2+gr^4/4)\de r}}=\frac{1}{\beta^{k}}
\frac{\int y^{2k} e^{-y^2/2}e^{-(g/\beta)y^4/4}\de y
}{\int e^{-y^2/2}e^{-(g/\beta)y^4/4}\de y}\sim\nonumber\\
&\sim& \frac{1}{\beta^{k}}
\frac{\int y^{2k} e^{-y^2/2}\de y
}{\int e^{-y^2/2}\de y}=\frac{(2k-1)!!}{\beta^{k}}, 
\end{eqnarray}
it follows from \eqref{epslim} and \eqref{alphalim} that $\varepsilon(\beta)\sim 1/\beta$  and 
$\alpha(\beta)\sim 3g$. As a consequence, the frequency renormalization coefficient
defined in \eqref{Omk}-\eqref{gamma}, turns out to be
\begin{equation}
\gamma(\varepsilon)=2\alpha(\beta)\langle u_2\rangle_\mu\sim
3g\varepsilon
\end{equation}
as $\beta\to\infty$, or $\varepsilon\to0$.
A significant question is then whether the renormalized frequency spectrum
\begin{equation}
\label{Omkloweps}
\Omega_k\sim\sqrt{1+3g\varepsilon}\ \omega_k\sim(1+3g\varepsilon/2)\omega_k
\end{equation}
and the the effective normal modes predicted by our mean field theory
coincide with those found in computing the Birkhoff normal form
of the FPUT Hamiltonian \eqref{H}. We show that the answer is affirmative: our statistical perturbation theory reduces to the Birhkoff one in the low $\varepsilon$ limit.

The Birkhoff normal form of the FPUT Hamiltonian \eqref{H} can be computed by first rewriting $H$ in terms of the complex Birkhoff variables 
\begin{equation}
z_k=\frac{\omega_kQ_k+\imath P_k}{\sqrt{2\omega_k}}\ \ ;\ \ 
z^*_k=\frac{\omega_kQ_k-\imath P_k}{\sqrt{2\omega_k}}\ ,
\end{equation}
where $Q_k$ and $P_k$ are defined in \eqref{QkPk} and $\imath$ is the imaginary unit.
The result is
\begin{equation}
H=\sum_{k=1}^{N-1}\omega_k|z_k|^2+gU_4(z,z^*)\ ,
\end{equation}
where $z=(z_1,\dots,z_{N-1})$ and $U_4(\bar z,z)$ has a certain explicit expression. The Birkhoff normal form, to leading order, is then given by replacing $U_4$ with its time-average $\overline{U}_4$ along the unperturbed flow of the quadratic Hamiltonian $H_2=\sum_k\omega_k|z_k|^2$. Such a flow is given by
\begin{equation}
\Phi^t(z,z^*)=(e^{-\imath\omega_1t}z_1,\dots,
e^{-\imath\omega_{N-1}t}z_{N-1},e^{\imath\omega_1t}z^*_1,\dots
e^{\imath\omega_{N-1}t}z^*_{N-1})\ ,
\end{equation}
and the time average $\overline{U}_4$ is defined by 
\begin{equation}
\overline{U}_4(z,z^*)=\lim_{T\to\infty}\frac{1}{T}\int_0^TU_4(\Phi^t(z,z^*))\de t\ .
\end{equation}
Observe that, of course, $\overline{H}_2=H_2$. We do not report the details of the long computation, that can be found in \cite{HeKa}. The final result is 
\begin{equation}
\label{Hbar}
\overline{H}=H_2+g\overline{U}_4=H_2+\frac{3g}{4N}H_2^2+R_N\ ,
\end{equation}
where the reminder term $R_N$ has a certain explicit expression, which is irrelevant in the limit of large $N$, since  $R_N=O(1)$ with respect to $N$. Neglecting the remainder $R_N$, and observing that 
$H_2=\sum_k\omega_k|z_k|^2=\frac{1}{2}\sum_k(P_k^2+\omega_k^2Q_k^2)$, the equations of motion associated to the Hamiltonian \eqref{Hbar} read
\begin{equation}
\label{Bireq}
\dot Q_k=\left(1+\frac{3g}{2}\frac{H_2}{N}\right)P_k\ \ ;\ \ 
\dot P_k=-\left(1+\frac{3g}{2}\frac{H_2}{N}\right)\omega_k^2Q_k\ .
\end{equation}
Now, observing that $H_2/N\sim \varepsilon$ if $\varepsilon=\overline{H}/N$ is small,
the equations \eqref{Bireq} in second order form read
\begin{equation}
\ddot Q_k=-\Omega_k^2Q_k\ ,
\end{equation}
with $\Omega_k$ given by \eqref{Omkloweps}. The above equation coincides with 
the effective normal mode equation \eqref{Qkmf2} in the limit of small $\varepsilon$. 

\section{Virial theorem and Kinchin ergodicity}
\setcounter{section}{3}
\label{Avirial}

The virial theorem for the FPUT system \eqref{H} is deduced by starting with the identity
\begin{equation}
\frac{\de}{\de t}\sum_nq_np_n=2K-2U_2-4gU_4\ .
\end{equation}
Now, both the time average (over infinite time) and the Gibbs mean value of  the time derivative on the left hand side vanish. Thus, considering first the Gibbsian mean 
$\langle\cdot\rangle_\mu$, one gets the virial identity
\begin{equation}
\label{virid}
\langle K\rangle_\mu-\langle U_2\rangle_\mu-2g\langle U_4\rangle_\mu=0\ .
\end{equation}
Taking the Gibbsian mean of the energy conservation law $K+U_2+gU_4=E$, and dividing both this equation and the virial identity \eqref{virid} by $N$, we get the virial system
\begin{equation}
\label{virsys}
\left\{\begin{array}{l} \langle k\rangle_\mu-\langle u_2\rangle_\mu-2g\langle u_4\rangle_\mu=0 \\
\langle k\rangle_\mu+\langle u_2\rangle_\mu+g\langle u_4\rangle_\mu=\varepsilon
\end{array}\right.\ ,
\end{equation}
where $k=K/N$ and $\varepsilon=H/N$. The second equation, i.e. the energy conservation law, coincides with our equation \eqref{epsbeta}, relating $\varepsilon$ to the inverse temperature $\beta$. By substituting 
$\langle u_4\rangle_\mu=\alpha\langle u_2^2\rangle_\mu$ in \eqref{virsys}, and taking into account that $\langle u_2^2\rangle_\mu\to\langle u_2\rangle_\mu^2$ as 
$N\to\infty$, which trivially follows from \eqref{uaav} and \eqref{ua2av}, we get a system in the two variables $\langle k\rangle_\mu$ and 
$\langle u_2\rangle_\mu$. In particular, the virial identity, first equation of \eqref{virsys}, gives
\begin{equation}
\label{virrel}
\langle k\rangle_\mu=\langle u_2\rangle_\mu+2g\alpha\langle u_2\rangle_\mu^2\ .
\end{equation}
Now, dividing the latter relation by $\langle u_2\rangle_\mu$ and remembering that 
$\langle k\rangle_\mu=1/(2\beta)$ and $\langle u_2\rangle_\mu\to\langle r^2\rangle_\phi/2$,
proves equation \eqref{virial}, i.e. the relation between our expression of $\gamma$ and that found by Lepri \cite{L98} with the Zwanzig-Mori method.  Another consequence of 
\eqref{virrel} is the following. Rewriting $k=K/N$ and $u_2=U_2/N$ in the normal mode variables \eqref{QkPk}, we find
\begin{equation}
\frac{\sum_k \langle P_k^2\rangle_\mu}{
\sum_k\omega_k^2\langle Q_k^2\rangle_\mu}=
\frac{\langle k\rangle_\mu}{\langle u_2\rangle_\mu}=1+2g\alpha\langle u_2\rangle_\mu=
1+\gamma\ ,
\end{equation}
which clearly suggests the frequency renormalization relation \eqref{Omk}.
 
In \cite{ACM95,AC01,AC02}, the authors started from the virial system \eqref{virsys}, but
with time averages in place of the ensemble ones, and their idea of "closure" was to replace 
$\overline u_4$ with 
$\alpha \overline u_2^2$ in order to solve the virial system for the two variables 
$\overline k$ and $\overline u_2$. Of course, in this way, the coefficient $\alpha$ can be determined only numerically. However, due to excellent agreement we found between time averaged quantities with ensemble averaged ones, as appears, for example, in Figure~\ref{fig:gamma}, we are forced to further explore the relation between the ensemble and time averages, in general. To such a purpose we make use of a fundamental idea of 
Kinchin \cite{K49}. Consider the specific potential energy $u_a$, defined in \eqref{ua},  and its time average over a finite time $T$, namely
\begin{equation}
\overline u_a(T)=\frac{1}{T}\int_0^Tu_a(s)\de s\ . 
\end{equation}
where $u_a(s)$ is $u_a$ along any motion. We also introduce the instantaneous fluctuation
$\delta u(s)=u_a(s)-\langle u_a\rangle_\mu$ and its time average 
$\overline{\delta u}(T)=\overline u(T)-\langle u_a\rangle_\mu$. Then,
one easily proves that
\begin{equation}
\label{Kinch}
\mu(\{|\overline{\delta u}(T)|>\eta\})\leq\frac{\langle (\delta u_a(0))^2\rangle_\mu}{\eta^2}=
O(1/N)\ .
\end{equation}
The last equality above has been proved previously, whereas the first inequality follows from the Chebyshev one followed by the Schwartz inequality in measure \cite{JP}. 
The estimate \eqref{Kinch} shows that one can reasonably substitute, when $N$ is large enough, time averages with ensemble ones, no matter how long the averaging time $T$ is (the last bound $O(1/N)$ in \eqref{Kinch} is independent of it; $T$ must only be large enough to ensure the validity of the virial identity). This is not due to ergodicity, but to the law of large numbers, which is the point of view of Kinchin \cite{K49}.

\end{document}